# Coupled Sublattice Melting and Charge-Order Transition in Two Dimensions

T.S. Smith,[1]† F. Ming,[1,2]† D. G. Trabada,[3] C. Gonzalez,[3] D. Soler-Polo,[3] F. Flores,[3] J. Ortega,[3] H.H. Weitering[1*]

[1] *Department of Physics and Astronomy, The University of Tennessee, Knoxville, TN 37996, USA.*

[2] *State Key Laboratory of Optoelectronic Materials and Technologies, School of Electronics and Information Technology, Sun Yat-sen University, Guangzhou 510275, China*

[3] *Departamento de Física Teórica de la Materia Condensada and Condensed Matter Physics Center (IFIMAC), Universidad Autónoma de Madrid, ES-28049 Madrid, Spain.*

\* Correspondence to: hanno@utk.edu.

† These authors contributed equally.

**Abstract:** Two-dimensional melting is one of the most fascinating and poorly understood phase transitions in nature. Theoretical investigations often point to a two-step melting scenario involving unbinding of topological defects at two distinct temperatures. Here we report on a novel melting transition of a charge-ordered K-Sn alloy monolayer on a silicon substrate. Melting starts with short-range positional fluctuations in the K sublattice while maintaining long-range order, followed by longer-range K diffusion over small domains, and ultimately resulting in a molten sublattice. Concomitantly, the charge-order of the Sn host lattice collapses in a multi-step process with both displacive and order-disorder transition characteristics. Our combined experimental and theoretical analysis provides a rare insight into the atomistic processes of a multi-step melting transition of a two-dimensional materials system.

**Main Text:** Low-dimensional solids are fundamentally different from three-dimensional (3D) bulk solids. This difference is especially manifest in phase transitions and critical phenomena as the balance between energy and entropy is profoundly altered in low dimension. The melting transition arguably provides the most striking contrast. 3D melting is known to be a first order (i.e. discontinuous) phase transition. In 2D, on the other hand, continuous melting transitions are possible and they are believed to be topological in nature (*1-5*). Most structural phase transitions in quasi 2D systems such as surfaces and interfaces can be described within Landau's conceptual framework of spontaneous symmetry breaking where the ground state of the system no longer possesses the full symmetry of the high-temperature phase (*6, 7*). Symmetry and dimensionality are the quintessential ingredients of this theoretical framework.

Charge-order transitions are especially ubiquitous in surface physics due to the strong coupling between the lattice and electronic degrees of freedom at the surface, as well as the extra freedom of the surface atoms to relax vertically. As a result, many surfaces 'reconstruct' (*8, 9*). While the precise driving force and mechanisms of these structural transitions are often under debate, the broken-symmetry ground state is characterized by a (partial) gap opening, which amounts to a lowering of the total electronic energy of the system (*8, 9*). The symmetry can be restored via thermal excitations of phonons, which in the strong coupling scenario can lead to an order-disorder transition at high temperature (*7, 10*). Below the transition temperature $T_C$, atoms oscillate around their equilibrium positions of the low temperature phase. Above $T_C$, the atoms fluctuate between different potential minima and the symmetry is fully restored via time- (or ensemble) averaging (*10*). A competing weak-coupling picture of the phase transition involves a phonon softening in the high temperature phase and a frozen phonon or lattice deformation below $T_C$ (*11*). Such a transition is called 'displacive' (*7*). Experimentally, these two scenarios are difficult to distinguish as many phase transitions exhibit a combination of weak coupling and strong coupling features (*12*).

Here, we report on a novel 2D transition in a charge-ordered K-Sn alloy monolayer on Si(111), whereby the 'melting' of the charge order in the Sn host lattice is preceded and driven by a true 2D melting transition in the alkali sublattice. Sublattice melting is a well-known phenomenon in bulk ionic conductors where an ionic insulator acquires a highly conductive state when the sublattice containing the lighter ions, typically an alkali or Ag ion, melts while the heavier sublattice or 'host' lattice remains intact (*13, 14*). A similar part-liquid part-crystalline state has been observed in thermoelectric compounds with strong chemical-bond hierarchy (*15*). Here we show that sublattice melting can also happen in 2D, and that it can have profound consequences for the charge-order phase transition in the host lattice. Specifically, the phase transition in the K-Sn adatom system turns out to be a multi-step process involving two intermediate regimes. The first one is characterized by short-range positional fluctuations in the alkali sublattice and partial band gap reduction of the charge-ordered condensate. The second intermediate regime involves longer-range diffusion of the K atoms, the nucleation of liquid-like islands, and concomitant melting of the charge-ordered state of the Sn host lattice. The latter involves both a displacive and order-disorder mechanism. The present study provides a rare glimpse into the atomistic processes of a two-dimensional melting transition, which remains one of the oldest outstanding problems in condensed matter physics (*4, 5*).

**Methods**.

Potassium was deposited from a getter source onto the Si(111)($\sqrt{3}\times\sqrt{3}$)$R30°$-Sn surface (henceforth Sn$\sqrt{3}$). The temperature of the Sn$\sqrt{3}$ substrate surface was kept below 120 K and the pressure was better than $2\times10^{-10}$ mbar. The Sn$\sqrt{3}$ reconstruction consists of a triangular array of Sn adatoms that are located at T4 adsorption sites of the Si(111) surface (*16-18*). The areal Sn coverage is 1/3 monolayer ($\Theta_{Sn} = 1/3$ML). Its preparation has been described in detail in Ref. 18. To ensure the consistency of the results, we repeated the experiments on different Si(111) substrates, including heavily-doped n-type substrates (As-doped; 0.002 Ωcm), moderately doped p-type substrates (B-doped; 0.02 Ωcm), and heavily-doped p-type substrates (B-doped; 0.001 Ωcm). We generally find that the n-type substrates produce fewer defects while only p-type substrates provide reliable spectroscopy data at low temperature (*19*). The surfaces were studied using scanning tunneling microscopy and spectroscopy (STM/STS). We used density functional theory (DFT) and DFT-based molecular dynamics (DFT-MD) simulations to study the ground state properties and atom dynamics of the K decorated Sn$\sqrt{3}$ system (see Supplementary Materials (SM) for details).

**Results and Discussion**.

Figure 1 shows STM images of a well-ordered ($2\sqrt{3}\times2\sqrt{3}$)$R30°$ structure (henceforth $2\sqrt{3}$-K) formed by deposition of K onto the Sn$\sqrt{3}$ surface at 77 K. The K coverage is $\Theta_K = 1/6$ ML and the unit cell has quadrupled relative to the Sn√3 substrate unit cell. See SM for the determination of the absolute K coverage. Here we show the result for the heavily-doped p-type substrate; the same structure is formed on the other substrates. The images reveal three distinct lattices, depending on the tunneling bias. The filled state image in Fig. 1A shows a Kagome lattice with three bright adatoms and one dim atom per unit cell. The dim atoms are located at the center of each hexagon. These are the Sn adatoms (see below). The center atoms can hardly be seen in this image but they do form a bright triangular lattice in the empty state image in Fig. 1B. The Kagome lattice is hardly visible at this bias (+ 1.0 V). The triangular lattice fades with increased bias and a bright honeycomb lattice appears at ~1.3 V as shown in Fig. 1C. These are the K atoms. They are located at the centers of the chained triangles that make up the Kagome lattice. STM imaging of this K honeycomb lattice becomes very problematic above 130 K indicating K motion on the surface.

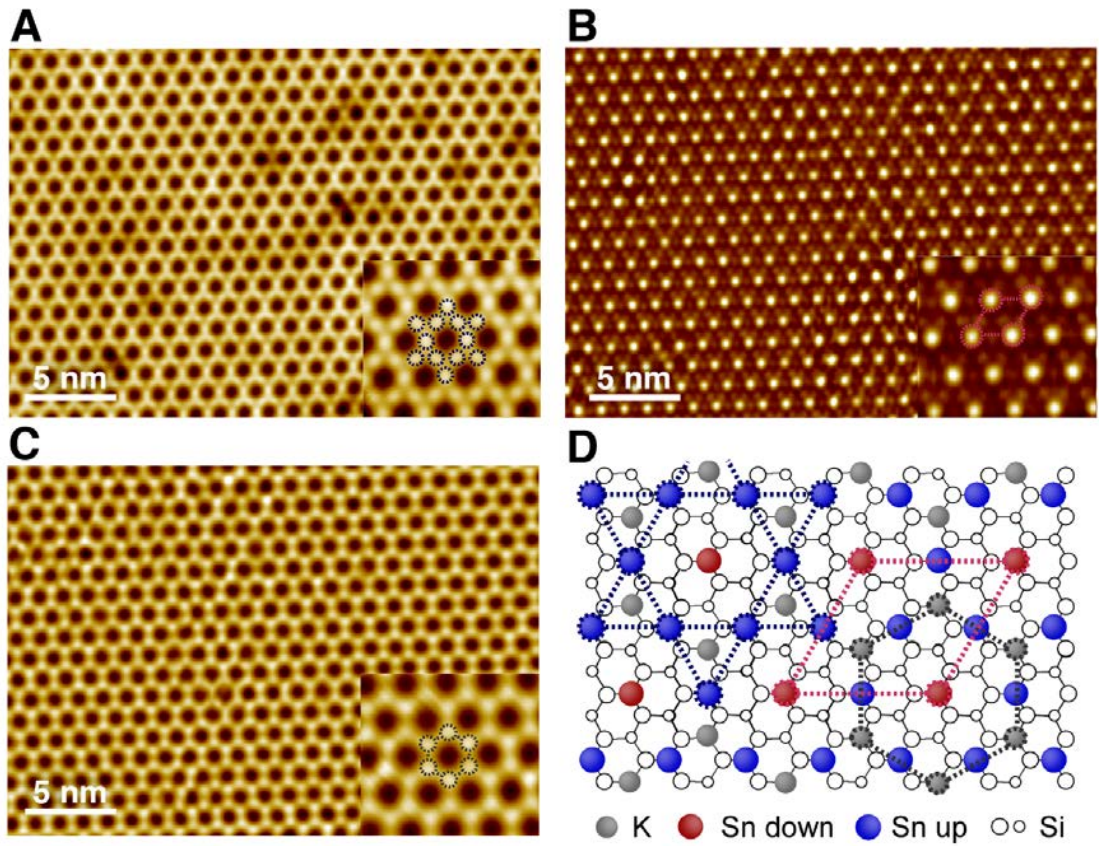

*Fig. 1.* **STM topography of the 2√3-K structure on heavily doped p-type Si(111).** *The images, recorded at 77 K, show the same surface area. The images are scanned at (A) -1.5 V and 1 nA, revealing a Kagome lattice comprised of the Sn up atoms; (B) 1.0 V and 1nA revealing the triangular lattice comprised of the Sn down atoms; and (C) 2.5 V and 1 nA, showing a honeycomb lattice corresponding to the K locations. The insets in (A-C) are zoomed-in images, which more clearly reveal the corresponding sublattices. (D) Atomic model of the 2√3-K structure on Si(111) (top view), obtained from the DFT calculations. The K atoms are located at T4 adsorption sites at the centers of the Sn 'up' triangles.*

Images for $\Theta_K \ll 1/6$ ML (Fig. S1) and $\Theta_K > 1/6$ ML (Fig. S2) reveal that the Kagome and triangular lattice sites in Figs. 1A and 1B coincide with the Sn adatom locations of the Sn√3 substrate. The registry of the Sn and K atoms from the STM images is illustrated in Fig. 1D. The images can easily be understood as follows. For $\Theta_K$ = 1/6 ML, there are four Sn adatoms and two K atoms per (2√3×2√3) unit cell. Assuming that the K atoms are ionized, there are six valence electrons to be distributed over four dangling bonds, resulting in three doubly-occupied dangling bond orbitals and one empty dangling bond orbital. The three adatoms with doubly occupied dangling bonds relax outward (*8, 10*), thus forming the (2√3×2√3) Kagome sublattice. Adatoms with empty dangling bonds relax inward, forming the (2√3×2√3) triangular sublattice (*8, 10*). This charge ordering phenomenon (*10, 20, 21*) is thus accompanied by a buckling distortion of the Sn adatom lattice (*10*). As we will show with STS, the ground state is insulating.

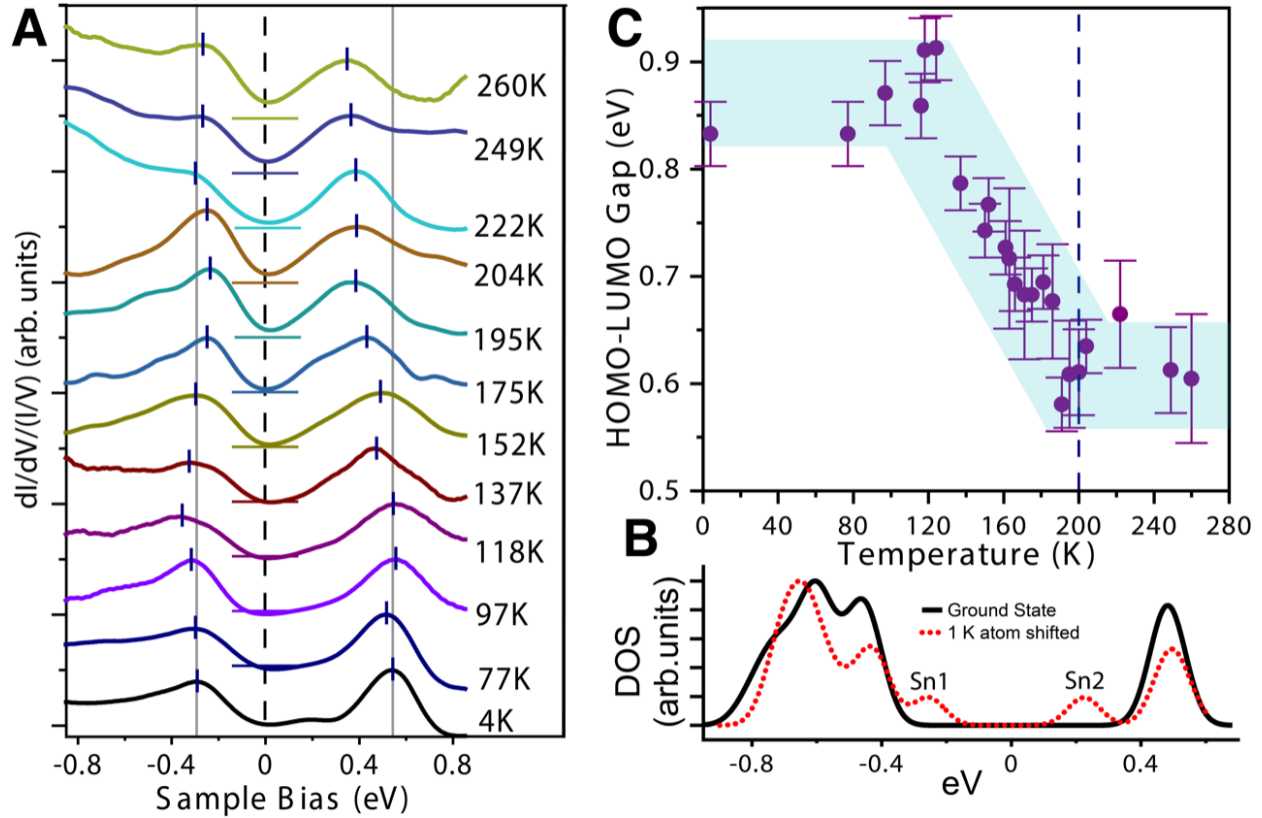

*Fig. 2. **Evolution of the HOMO-LUMO states**. (A) Temperature dependent dI/dV spectra of the 2√3-K structure on the moderately doped p-type substrate. The maxima of the HOMO and LUMO peaks are indicated by vertical tick marks. Grey vertical lines mark the HOMO/LUMO locations at 4K. The reduction of the HOMO-LUMO splitting above 118K is primarily attributed to the downward shift of the LUMO state. Horizontal tick marks below each spectrum indicate the zero of the tunneling conductance. The increase in zero-bias conductance (ZBC) above 175K, signals a gradual insulator-metal transition. (B) Theoretical LDOS of the (2√3×2√3)R30° ground state and (4√3×4√3)R30° excited-state system with one displaced potassium atom. Note that the potassium displacement produces new band gap states, labeled Sn1 and Sn2, at about -0.25 and + 0.22 eV, thus reducing the band gap. (C) Evolution of the experimental HOMO-LUMO gap, as measured from the positions of the LDOS maxima marked by the vertical tick marks in (A).*

This scenario is fully confirmed with plane-wave DFT calculations using the *Quantum Espresso* code (*22*) (for details, see SM). The total-energy-minimized structure indicates a 0.23 Å height difference between the Sn up atoms and Sn down atoms or, equivalently, between the Kagome and triangular sublattices, respectively; the K atoms are 0.75 Å above the Sn up atoms. Furthermore, STM image simulations (Fig. S9 and S10) are fully consistent with the experimental images in Fig. 1.

Figure 2A shows the temperature-dependent normalized d$I$/d$V$ spectra of the 2√3-K structure on the moderately doped p-type substrate, reflecting the local density of states (LDOS) (*23*). The spectra in Figure 2B were calculated with DFT, using the PBE0 hybrid functional (*24, 25*). The spectrum of the ground state system, indicated by the solid black line is in good agreement with the spectrum measured at 4 K, although the use of the hybrid functional slightly overestimates the band gap. The experimental spectrum is characterized by a DOS peak centered at about -0.3 eV, and a DOS peak centered at about +0.5 eV. The Highest Occupied Molecular Orbital (HOMO) at about -0.3 eV is associated with the dangling bond orbitals of the Sn up atoms, as the corresponding STM images show the Kagome lattice (Fig. 1A). The Lowest Unoccupied Molecular Orbital (LUMO) at about +0.5 V is associated with the Sn down atoms of the triangular lattice (see the projected DOS shown in Fig. S10). The HOMO-LUMO splitting is a measure of the Sn up-down lattice distortion in the charge ordered 2√3-K phase. As such, its magnitude can be considered a *bona-fide* order parameter for the charge-ordered condensate (*7*). Its temperature dependence is shown in Fig. 2C. This figure reveals a gradual decrease of the order parameter, starting at 130 K and leveling off at 200 K where it reduces to approximately 70 percent of its original value. Interestingly, the HOMO-LUMO splitting remains constant above 200 K. The order parameter decrease should be primarily attributed to the downward shift of the LUMO state, or more precisely, to the development of a shoulder on the low energy side of the LUMO state (see SM). This shoulder is clearly visible in the 175 K spectrum and appears to be centered near +0.2 eV. This broadening leads to a gradual insulator-metal transition near 175 K. As will be discussed later, the experimental results together with the DFT calculations indicate that the broadening and shift of the LUMO is related to short-range positional fluctuations of the potassium atoms while the system maintains its long-range (2√3x2√3) translational order.

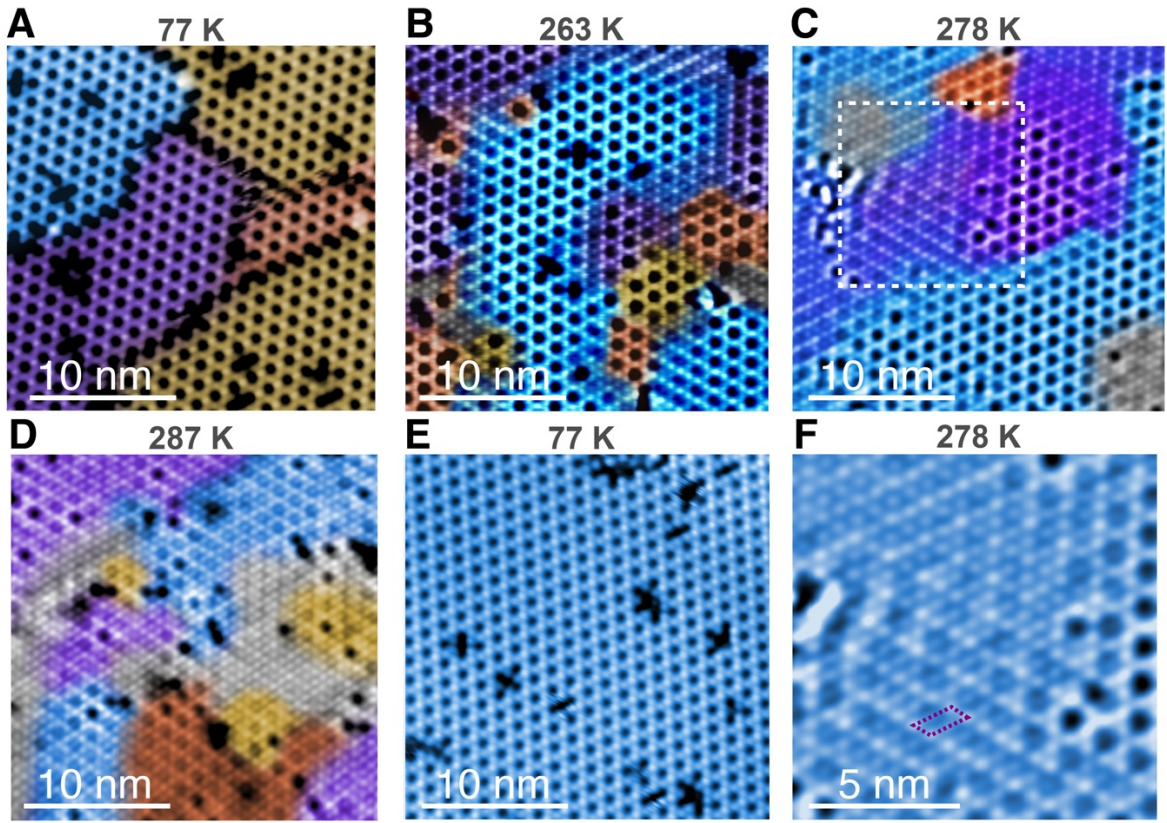

*Fig. 3. **Temperature dependent topography of the 2√3-K structure on the n-type substrate**. The sample was grown at 120 K, cooled to and imaged at 77 K (**A**). It was then gradually warmed to room temperature (**B-D**), and finally cooled back down to 77 K (**E**). Four out-of-phase domains of the 2√3-K structure are displayed in different colors; a different (√3x√3)R30° phase that appears only above 200 K is colored in gray. A region showing (√3x2√3) ordering in (C) is marked with a dashed white square and its enlarged image is shown in (F), where the (√3x2√3) unit cell is marked by a dashed rhomboid. All images were obtained with a sample bias of -1.5V and 30 pA current.*

Figure 3 shows the temperature dependent topography of the 2√3-K structure on the n-type substrate (similar behavior is seen on other substrates, see Figs. S3 and S4). At 77 K, the surface is fully covered with the 2√3-K structure. Since there are four equivalent adatom sites per (2√3×2√3) unit cell prior to K deposition, there are four different domains, indicated with different colors. By slowly warming the sample up to room temperature, small patches of the 2√3-K phases convert to a new (√3×√3) lattice, as seen in the areas marked in gray in Fig. 3B-D. Some of the areas are not fully converted to a (√3×√3) lattice and still show weak (2√3×2√3) ordering around the defects (Fig. 3D) or a stripe-like (√3×2√3) structure (Fig. 3F). Subsequent cool down to 77 K produces a single domain 2√3-K lattice (Fig. 3E), in contrast to the high density of domain boundaries of 2√3-K phase in Fig. 3A or the defect pined 2√3-K patches in Fig. 3D. This indicates that K atoms diffuse over long distances during the annealing cycle and that defects do not play a significant role in the nucleation of the 2√3-K phase.

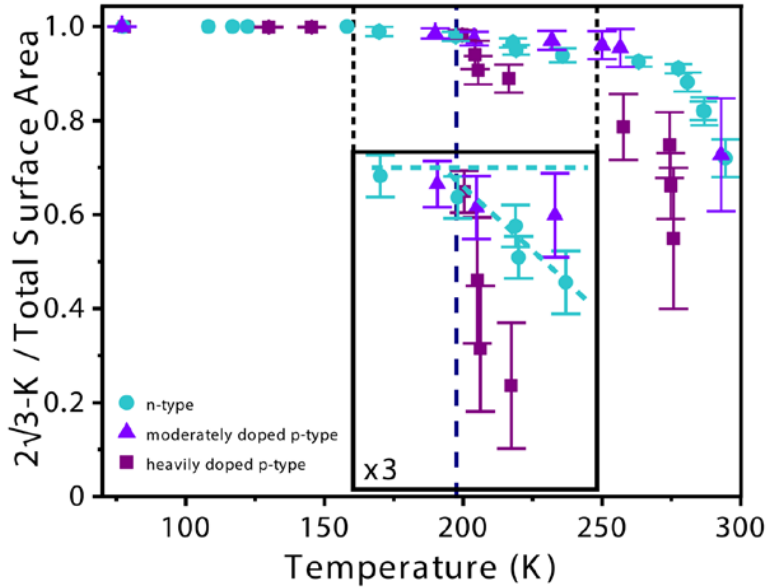

*Fig. 4. **Area fraction of the low-temperature 2√3-K phase on all three substrates**. Dashed lines indicate the location of the kink that occurs upon the appearance of the √3×√3 phase. The inset shows that this occurs on all substrates at the same temperature, albeit with a different rate on different doping types.*

Figure 4 shows the area fraction of the 2√3-K phase as a function of temperature for all three substrates. Below 200 K, the 2√3-K phase covers the entire surface. Above 200 K, the (2√3×2√3), (2√3×√3), and (√3×√3) structures all coexist. The onset temperature for the phase coexistence coincides with the end point of the transition seen in Fig. 2C. This can be seen from the zoomed-in data in the inset, most noticeably for the heavily-doped p-type silicon substrates, showing a clear reduction in the area fraction of the 2√3-K phase starting at 200 K. While this reduction is slow initially, the area fraction of the 2√3-K phase drops very steeply near 275 K, possibly indicating a first order phase transition to the (√3×√3) phase. We tentatively conclude that there are at least two transitions. One is a continuous transition, characterized by short-range positional fluctuations of the akali atoms, from the long-range-ordered 2√3-K phase below 130K to a quasi-ordered phase above 200 K. The other is the complete melting of the charge order near 275 K (See Fig. S5).

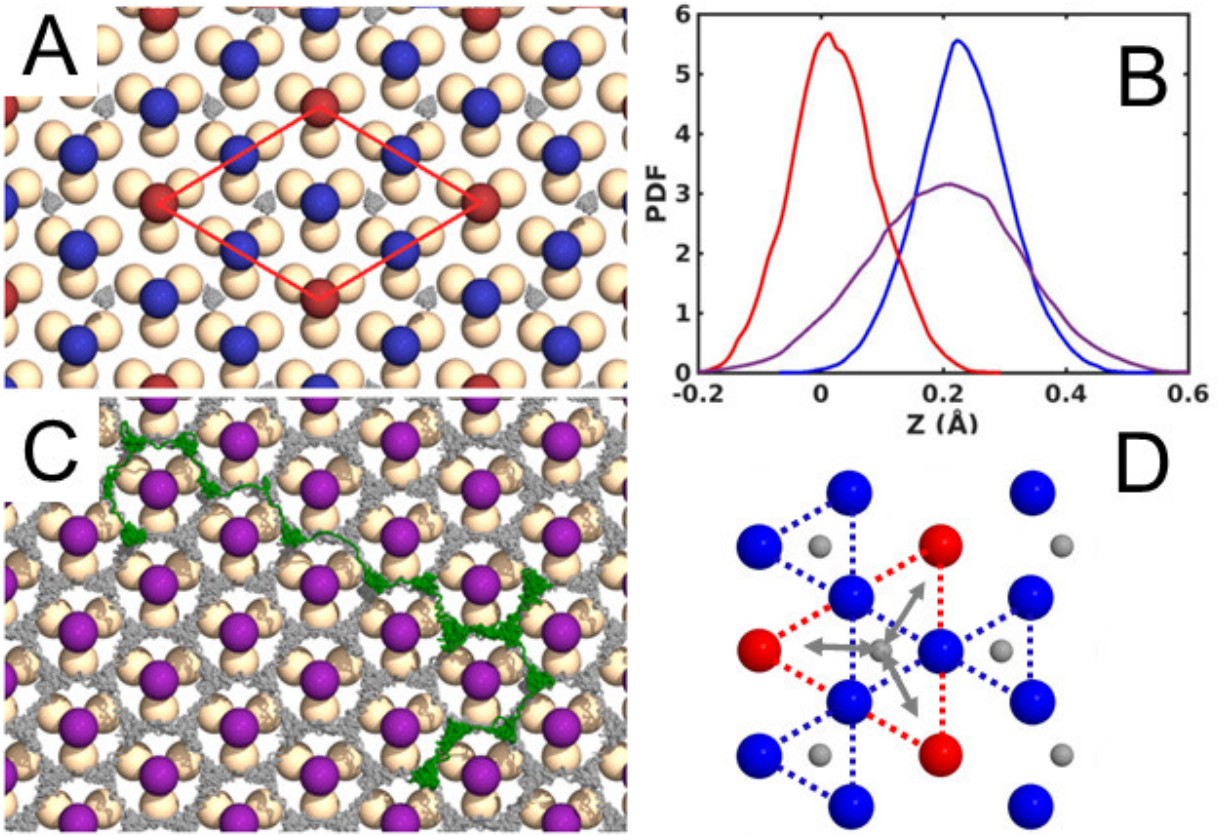

*Fig. 5. **Atomic dynamics in the K/Sn layer**. (**A, C, D**) Top view of the 2√3-K surface. Gray lines follow the motion of the potassium atoms at 77 K (A) and 300 K (C). The green line in (C) follows the trajectory of a single K atom along a 300 ps DFT-MD simulation. Blue and red marbles represent the Sn-up and Sn-down atoms, respectively. At 300 K all Sn adatoms are equivalent (purple marbles). Beige marbles represent the Si atoms below. (**B**) Probability Distribution Functions for the Sn-up (blue) and Sn-down (red) adatom heights at 77K, and for the Sn atoms at 300 K (purple); z = 0.0 Å is the height of the Sn-down atoms in the 2√3-K structure. (D) Schematic illustration of the short-range positional fluctuations of the K atoms to nearest neighbor T4 sites.*

To understand the atomic mechanisms of this phase transition, we have performed very long DFT-MD simulations at various temperatures using the *Fireball* code (*26*) (see SM for details). In these simulations we have used a basis set of numerical atomic-like orbitals that yields a good description for the K/Sn/Si(111) system (see SM). The DFT-MD simulations are performed for a periodic system with a (2√3×2√3) unit cell and thus cannot account for the complex domain structures in Fig. 3. However, they do provide insights into the K distribution and associated Sn adatom relaxations within each of these domains, thus elucidating the interplay between the melting of the alkali sublattice and the charge order transition.

At 77 K, the system exhibits long-range (2√3×2√3) order. The DFT-MD simulations show K atoms 'rattling' around their stable positions at the T4 sites inside the Sn adatom triangles that make up the Kagome lattice, see Figs. 1D and 5A. Meanwhile, the Sn atoms oscillate around their up or down equilibrium positions (Fig. 5B). Note that in this configuration, the Sn up-atoms have two K nearest neighbors, while the Sn down-atoms have none.

At higher temperatures (*e.g.* see Fig. 5C), alkali atoms start jumping to neighboring potential wells defined by the other T4 sites on the Si(111) surface, excepting those already occupied by Sn adatoms, see Figs. 5C, 5D and S7. In total, there are eight T4 sites per (2√3×2√3) unit cell available to the alkali atoms (Fig. 1D). Accordingly, there are 28 possible configurations in which the two K atoms in the unit cell occupy two of these potential wells. Four of these K-configurations correspond to the 2√3-K domains that appear in Fig. 3A, which according to the STM results remains the dominant structure up to at least 200 K.

To better understand the energetics of these positional fluctuations and the concomitant changes in the electronic structure, we performed plane-wave DFT calculations for a large (4√3×4√3) unit cell in which one K atom is displaced laterally towards a neighboring T4 site (Fig. 5D). In this configuration, which can be viewed as a snapshot of the K positional fluctuations, the K atom moves away from one Sn-up atom (say $Sn_1$) of the Kagome lattice and comes close to one Sn-down atom (say $Sn_2$) of the triangular lattice (Fig. S7). This induces a downward displacement ($\Delta z \approx 0.04$ Å) of $Sn_1$ and an upward displacement ($\Delta z \approx 0.07$ Å) of $Sn_2$; the dangling bond states of these two Sn atoms move into the gap, as indicated by the red dotted spectrum in Fig. 2B, thus resulting in the reduction of the HOMO-LUMO gap. Although the $Sn_1$ and $Sn_2$ atoms have one K nearest neighbor, the different environment and adatom heights of $Sn_1$ and $Sn_2$ ($\Delta z \approx 0.11$ Å) produce a small but finite gap.

This is in good agreement with the evolution of the HOMO-LUMO gap shown in Fig. 2C. In particular, the downshift of the LUMO in Fig. 2A can be largely attributed to the emergence of a shoulder on the low energy side of the LUMO. This is clearly evident from a fitting of the experimental LUMO spectra (not shown) and from an eyeball inspection of the 175 K spectrum. The spectral weight transfer from the original HOMO and LUMO states to the new $Sn_1$ and $Sn_2$ states (Fig. 2B) should indeed be most noticeable on the LUMO side where one out of four dangling bond states per (4√3×4√3) unit cell shifts to lower energy forming the $Sn_2$ state. On the HOMO side, only one out of twelve dangling bond states shifts to higher energy forming the $Sn_1$ state. As the number of short range hopping events increases with temperature, the LUMO shoulder should become more pronounced, consistent with the observation (Fig. 2C).

We have also estimated the energy barriers for K jumps to neighboring T4 positions, using this (4√3×4√3) unit cell; the barrier for a K short range positional fluctuation is 0.16 eV; once there, the energy barrier to return to the original position is only 0.09 eV while the barrier to further move to another T4 site is 0.20 eV. Thus, at *e.g.* T = 150 K it is much more likely that the shifted K atom returns to its original site, hence confirming the notion that the K fluctuations at 150 K are short ranged. This is consistent with the time-averaged STM measurements, showing that the system still remains in the 2√3-K configuration below 200 K.

In the DFT-MD simulations at 300 K (Fig. 5C) the K sublattice is molten and the alkali atoms visit all the possible K-configurations (*27*). In the predominant configurations in these dynamics the Sn adatoms align in (√3×2√3) rows (see Figs. 3F and S6); the height of the Sn atoms in the upper row is the same as that of the up atoms of the 2√3-K phase while the Sn atoms in the lower row are only ~ 0.1 Å below (Fig. S6). Hence, the buckling distortion of the charge-ordered 2√3-K phase is greatly reduced due to the upward relaxation of the down-atoms of the triangular Sn sublattice. This can be viewed as a displacive transition: at low T the Sn up and down atoms oscillate around their equilibrium positions, z ~ 0.23 Å and z ~ 0.00 Å, respectively, while at high T the Sn atoms oscillate around a new vertical position, z ~ 0.20 Å (see Figs. S6, S8). Since at 300 K the Sn adatoms fluctuate between the different (√3×2√3) configurations induced by the alkali atoms, the distinction between Sn up and down is totally lost, see Figs. 5B and S8, and the system ultimately transforms into a (√3×√3) phase with a total collapse of the charge ordering. This final step is typical of an order-disorder transition where the disorder refers to dynamical fluctuations between different (√3×2√3) structures (*10, 28, 29*).

The following picture emerges. The melting of the 2√3-K charge ordered condensate starts at about 130 K and proceeds through two intermediate regimes. The first one is characterized by short-range positional fluctuations of the K atoms, as illustrated in Fig. 5D. These are mostly nearest neighbor hopping events where the K atoms always return to their original sites, thus maintaining long-range (2√3×2√3) order up to 200 K. These fluctuations are accompanied by up-down relaxations of the neighboring Sn atoms and become more prominent with increased temperature, inducing the gradual reduction of the HOMO-LUMO gap.

The second intermediate regime starts at 200 K, the nucleation temperature of the (√3×2√3) and (√3×√3) domains, which must be preceded by significant short-range fluctuations described above. In this regime, K atoms diffuse over longer distances. The formation of the (√3×2√3) domains within the Sn host lattice involves a displacive mechanism while the formation of the (√3×√3) domains involves an additional order-disorder transition (in similarity to the high T phase). As shown in Fig. 4, the long-range translational order of the 2√3-K phase now begins to diminish gradually as the pre-melted (√3×2√3) domains and molten (√3×√3) phase become more dominant. Coexistence between the 2√3-K, (√3×2√3) and (√3×√3) structures is typical of a pre-empted first order melting transition on a rigid substrate, where solid and liquid phases can coexist over a wide temperature range, due to long-range elastic interactions between the various domains (*30*).

The present scenario is somewhat reminiscent of the Kosterlitz, Thouless, Halperin, Nelson, and Young (KTHNY) theory for 2D melting (*1-3*). In this theory, 2D melting proceeds via an intermediate 'hexatic' phase, which is characterized by exponential decay of positional order correlations and power law decay of orientational correlations (quasi–long-range order). Recent studies employing advanced simulations on simplified model systems suggest that the solid-hexatic transition is continuous while the hexatic liquid transition may be discontinuous (i.e., first order) (*4*). Both transitions arise from the subsequent unbinding of topological defects such as paired dislocations and formation of free disclinations. The current situation appears to be quite a bit more complicated due to the presence of a substrate, absence of six-fold rotational order, and the coupling with the charge-order transition. Yet the microscopic multi-step melting scenario discussed here presents interesting similarities (and differences) with the two-step continuous-discontinuous melting transition mentioned above (*4*). Whether or not the coexisting periodicities and domain walls are topologically entangled (*31*) near the phase transition temperature and/or whether topological excitations play a role in this novel melting phenomenon remains to be determined.

**References and Notes:**

1. J. M. Kosterlitz, D. J. Thouless, Ordering, metastability and phase transitions in two-dimensional systems. J. Phys. C: Solid State Phys. **6**, 1181 (1973).
2. B. I Halperin, D. R. Nelson, Theory of two-dimensional melting. Phys. Rev. Lett. **41**, 121 (1978).
3. A. P. Young, Melting and the vector Coulomb gas in two dimensions. Phys. Rev. B **19**, 1855 (1979).
4. E. P. Bernard, W. Krauth, Two-step melting in two dimensions: first-order liquid-hexatic transition. Phys. Rev. Lett. **107**, 155704 (2011).
5. V. N. Ryzhov, E. E. Tareyeva, Yu D. Fomin, and E. N. Tsiok, Berezinskii–Kosterlitz–Thouless transition and two-dimensional melting. Phys.-Usp. **60**, 857-885 (2017).
6. L. D. Landau, E. M. Lifshitz, Statistical Physics (Pergamon Press Ltd, New York, ed. 3, 1980).
7. R.M. White, T.H. Geballe, Long-Range Order in Solids (Academic Press (1979).
8. M.-C. Desjonqueres, D. Spanjaard, Concepts in Surface Physics (Springer-Verlag, Berlin, 1993).
9. C.B. Duke, Semiconductor Surface Reconstruction: The structural chemistry of two-dimensional surface compounds. Chem. Rev. **96**, 1237-1260 (1996).
10. J. Avila, A. Mascaraque, E. G. Michel, M. C. Asensio, G. Le Lay, J. Ortega, R. Pérez, F. Flores, Dynamical fluctuations as the origin of a surface phase transition in Sn/Ge(111). Phys. Rev. Lett. **82**, 442 (1999).
11. W. Kohn, Image of the Fermi surface in the vibration spectrum of a metal. Phys. Rev. Lett. **2**, 393 (1959).
12. T. Aruga, Peierls transition on Cu(001) covered with heavier p-block metals. Surface Surf. Sci. Rep. **61**, 283-302 (2006).
13. V. Naden Robinson, H. Zong, G. J. Ackland, G. Woolman, A. Hermann, On the chain-melted phase of matter. Proc. Natl. Acad. Sci, **116**, 10297-10302 (2019)
14. J.D. Fan, R. Reiter, S.C. Moss, Dynamics of a 2D liquid on a periodic substrate: Rb in graphite. Phys. Rev. Lett. **64**, 188 (1990).
15. W. Qiu, L. Xi, P. Wei, X. Ke, J. Yang, W. Zhang, Part-crystalline part-liquid state and rattling-like thermal damping in materials with chemical-bond hierarchy. Proc. Natl. Acad. Sci., **111**, 15031-15035 (2014).
16. S. Modesti, L. Petaccia, G. Ceballos, I. Vobornik, G. Panaccione, G. Rossi, L. Ottaviano, R. Larciprete, S. Lizzit, A. Goldoni, Insulating ground state of Sn/Si(111)-(√3×√3)R30°. Phys. Rev. Lett. **98**, 126401 (2007).

17. G. Li, P. Höpfner, J. Schäfer, C. Blumenstein, S. Meyer, A. Bostwick, E. Rotenberg, R. Claessen, W. Hanke, Magnetic order in a frustrated two-dimensional atom lattice at a semiconductor surface. Nature Comm **4**, 1620 (2013).

18. F. Ming, S. Johnston, D. Mulugeta, T. S. Smith, P. Vilmercati, G. Lee, T. A. Maier, P. C. Snijders, H. H. Weitering, Realization of a hole-doped Mott insulator on a triangular silicon lattice. Phys. Rev. Lett. **119**, 266802 (2017).

19. F. Ming, D. Mulugeta, W. Tu, T. S. Smith, P. Vilmercati, G. Lee, Y.-T. Huang, R. D. Diehl, P. C. Snijders, H. H. Weitering, Hidden phase in a two-dimensional Sn layer stabilized by modulation hole doping. Nature Communications **8**, 473 (2017).

20. J. M. Carpinelli, H. H. Weitering, E. W. Plummer, R. Stumpf, Direct observation of a surface charge density wave. Nature **381**, 398-400 (1996).

21. J. M. Carpinelli, H. H. Weitering, M. Bartkowiak, R. Stumpf, E. W. Plummer, Surface Charge Ordering Transition: α Phase of Sn/Ge(111). Phys. Rev. Lett. **79**, 2859 (1997).

22. P. Giannozzi, S. Baroni, N. Bonini, M. Calandra, R. Car, C. Cavazzoni, D. Ceresoli, G. L Chiarotti, M. Cococcioni, I. Dabo, A. Dal Corso, S. de Gironcoli, S. Fabris, G. Fratesi, R. Gebauer, U. Gerstmann, C. Gougoussis, A. Kokalj, M. Lazzeri, L. Martin-Samos, N. Marzari, F. Mauri, R. Mazzarello, S. Paolini, A. Pasquarello, L. Paulatto, C. Sbraccia, S. Scandolo, G. Sclauzero, A. P Seitsonen, A. Smogunov, P. Umari, R. M Wentzcovitch, Quantum Espresso: A modular and open-source software project for quantum simulations of materials. J. Phys. Condens. Matter **21**, 395502 (2009).

23. J. A. Stroscio, R. M. Feenstra, A. P. Fein, Electronic structure of the Si(111)2 × 1 surface by scanning-tunneling microscopy. Phys. Rev. Lett. **57**, 2579 (1986).

24. J. P. Perdew, M. Ernzerhof, K. Burke, Rationale for mixing exact exchange with density functional approximations. J. Chem. Phys. **105**, 9982 (1996).

25. C. Adamo, V. Barone, Physically motivated density functionals with improved performances: The modified Perdew–Burke–Ernzerhof model. J. Chem. Phys. **110**, 6158 (1999).

26. J. P. Lewis, P. Jelínek, J. Ortega, A. A. Demkov, D. G. Trabada, B. Haycock, H. Wang, G. Adams, J.K. Tomfohr, E. Abad, H. Wang, D. A. Drabold, Advances and applications in the FIREBALL *ab initio* tight-binding molecular-dynamics formalism. Phys. Status Solidi B **248**, 1989-2007 (2011).

27. Since these are periodic DFT-MD simulations, when one K atom is leaving the (2√3×2√3) unit cell in one side, another K atom is entering the unit cell on the opposite side. In Figs. 5A and 5C we show the motion of the K atoms of the central and surrounding unit cells.

28. C. González, F. Flores, and J. Ortega, Soft phonon, dynamical fluctuations, and a reversible phase transition: indium chains on silicon. Phys. Rev. Lett. **96**, 136101 (2006).

29. D. G. Trabada, J. I. Mendieta-Moreno, D. Soler-Polo, F. Flores, J. Ortega, DFT molecular dynamics and free energy analysis of a charge density wave surface system. Appl. Surf. Sci. **479**, 260-264 (2019).

30. J. B. Hannon, F.-J. Meyer zu Heringdorf, J. Tersoff, R. M. Tromp, Phase coexistence during surface phase transitions. Phys. Rev. Lett. **86**, 4871 (2001).

31. G. Gye, E. Oh, H.W. Yeom, Topological landscape of competing charge density waves in 2$H$−NbSe$_2$. Phys. Rev. Lett. **122**, 016403 (2019).

32. S. Yi, F. Ming, Y.-T. Huang, T. S. Smith, X. Peng, W. Tu, D. Mulugeta, R. D. Diehl, P. C. Snijders, J.-H. Cho, and H. H. Weitering, Atomic and electronic structure of doped Si(111)(2√3×2√3)R30°-Sn interfaces. Phys. Rev. B **97**, 195402 (2018)

33. L. Chaput, C. Tournier-Colletta, L. Cardenas, A. Tejeda, B. Kierren, D. Malterre, Y. Fagot-Revurat, P. Le Fèvre, F. Bertran, A. Taleb-Ibrahimi, D. G. Trabada, J. Ortega, and F. Flores, Giant alkali-metal-induced lattice relaxation as the driving force of the insulating phase of alkali-Metal/Si(111):B. Phys. Rev. Lett. **107**, 187603 (2011)

34. J. P. Perdew, K. Burke, and M. Ernzerhof, Generalized Gradient Approximation made simple. Phys. Rev. Lett. **77**, 3865 (1996).

35. S. Goedecker, M. Teter and J. Hutter, Separable dual-space Gaussian pseudopotentials. Phys. Rev. B **54**, 1703 (1996).

36. C. Hartwigsen, S.Goedecker, and J. Hutter, Relativistic separable dual-space Gaussian pseudopotentials from H to Rn. Phys. Rev. B **58**, 3641 (1998).

37. M. J. Monkhorst and J. D. Pack, Special points for Brillouin-zone integrations, Phys. Rev. B **13**, 5188 (1976).

38. A. D. Becke, Density-functional exchange-energy approximation with correct asymptotic behavior. Physical Review A **38**, 3098–3100 (1988).

39. C. Lee, W. Yang, R. G. Parr, Development of the Colle-Salvetti correlation-energy formula into a functional of the electron density. Physical Review B **37**, 785 (1988).

40. M. A. Basanta, Y. J. Dappe, P. Jelínek, J. Ortega, Optimized atomic-like orbitals for first-principles tight-binding molecular dynamics. Comput. Mater. Sci. **39**, 759 (2007).

41. J. M. Blanco, C. González, P. Jelínek, J. Ortega, F. Flores, R. Pérez, First-principles simulations of STM images: From tunneling to the contact regime. Phys. Rev. B **70**, 085405 (2004).

42. C. González, P. C. Snijders, J. Ortega, R. Pérez, F. Flores, S. Rogge, H. H.Weitering, Formation of Atom Wires on Vicinal Silicon. Phys. Rev. Lett. **93**, 126106 (2004).

**Acknowledgments:** We thank P. Vilmercati for technical support. **Funding:** The experimental work was funded by the National Science Foundation under Grant No. DMR 1410265. TSS acknowledges the Center for Materials Processing, a Tennessee Higher Education Commission supported Accomplished Center of Excellence, for partial financial support. The theoretical work was funded by MINECO under the projects MAT2017-88258-R and MDM-2014-0377 (María de Maeztu Programme for Units of Excellence in R&D). The authors acknowledge the computer resources at Cibeles and the technical support provided by the Scientific Computing Center at UAM, project FI-2019-0028. **Author contributions:** T.S.S. and F.M. conducted the experimental investigations and data analysis and contributed to the manuscript preparation. D.G.T. and C.G. conducted the theoretical analysis, wrote the software and contributed to the review and editing. D.S.-P wrote the software and contributed to the theoretical analysis. F.F. contributed to the conceptualization, review and editing. J.O. and H.H.W. wrote the original draft, contributed to the conceptualization, review and editing, and supervised the project. **Competing interests:** Authors declare no competing interests. **Data and materials availability:** all data is available in the manuscript or the supplementary materials.

**Supplementary Materials:**

Supplementary Text

Figs. S1 to S8

Tables S1

References (*32–42*)